\documentclass[prl, preprintnumbers ,showpacs,amsmath,amssymb, superscriptaddress,twocolumn]{revtex4}
\usepackage{bm}
\usepackage{amsmath}
\usepackage{amssymb}
\usepackage{amsthm}
\usepackage{amsfonts}
\usepackage{enumerate}
\usepackage{latexsym}
\usepackage{times}
\usepackage{graphicx}
\usepackage{color}
\usepackage{makeidx}
\usepackage{ifpdf}

\begin{document}

\title{Tuning the competition between ferromagnetism and antiferromagnetism in a half-doped manganite through magnetoelectric coupling}

\author{Di~Yi} \email{yid@berkeley.edu}
\affiliation{Department of Materials Science and Engineering and Department of Physics, University of California, Berkeley, CA 94720, USA}
\author{Jian~Liu} \email{jian.liu@berkeley.edu}
\affiliation{Department of Materials Science and Engineering and Department of Physics, University of California, Berkeley, CA 94720, USA}
\affiliation{Materials Sciences Division, Lawrence Berkeley National Laboratory, Berkeley, CA 94720, USA}
\author{Satoshi~Okamoto}
\affiliation{Materials Science and Technology Division, Oak Ridge National Laboratory, Oak Ridge, TN, 37831, USA}
\author{Suresha~Jagannatha}
\affiliation{National Center for Electron Microscopy, Lawrence Berkeley National Laboratory, Berkeley, CA 94720}
\author{Yi-Chun~Chen}
\affiliation{Department of Physics, National Cheng Kung University, Taiwan 701}
\author{Pu~Yu}
\affiliation{State Key Laboratory for Low-Dimensional Quantum Physics, Department of Physics, Tsinghua University, Beijing 100084, PRC}
\author{Ying-Hao~Chu}
\affiliation{Department of Materials Science and Engineering, National Chiao Tung University, Hsinchu, Taiwan 300109}
\author{Elke~Arenholz}
\affiliation{Advanced Light Source, Lawrence Berkeley National Laboratory, Berkeley, CA 94720, USA}
\author{R.~Ramesh}
\affiliation{Department of Materials Science and Engineering and Department of Physics, University of California, Berkeley, CA 94720, USA}
\affiliation{Current address: Oak Ridge National Laboratory, Oak Ridge, TN 37831}

\date{\today}

\begin{abstract}
  We investigate the possibility of controlling the magnetic phase transition of the heterointerface between a half-doped manganite La$_{0.5}$Ca$_{0.5}$MnO$_{3}$ and a multiferroic BiFeO$_{3}$ through magnetoelectric coupling. Using macroscopic magnetometry and element-selective x-ray magnetic circular dichroism at the Mn and Fe \emph{L}-edges, we discover that the ferroelectric polarization of BFO controls simultaneously the magnetization of BFO and LCMO. X-ray absorption spectra at oxygen \emph{K}-edge and linear dichroism at Mn \emph{L}-edge suggest that the interfacial coupling is mainly derived from the superexchange between Mn and Fe \emph{t}$_{2g}$ spins. The combination of x-ray absorption spectroscopy and mean field theory calculations reveals that the \emph{d}-electron modulation of Mn cations changes the magnetic coupling in LCMO, which controls the enhanced canted moments of interfacial BFO via the interfacial coupling. Our results demonstrate that the competition between ferromagnetic and antiferromagnetic instability can be modulated by an electric field at the heterointerface, providing another pathway for the electrical field control of magnetism.
\end{abstract}

\maketitle

Over the past few years, complex oxide heterointerfaces have been extensively studied due to the novel phenomena that emerge at such interfaces and differ from the individual bulk components of the heterostructure \cite{Hwang,Zubko,Chakhalian}. Particularly, there has been a burst of activity to understand the intriguing interfacial magnetoelectric coupling \cite{Eerenstein,Cheong,Ramesh,Fiebig} in heterostructures consisting of ferroelectric (FE)/multiferroic insulator and ferromagnetic (FM) metal. By controlling the FE polarization and the underlying charge degree of freedom, one can manipulate the spin and orbital degrees of freedom \cite{Vaz,Burton,Wu0,Wu1}, and achieve magnetoelectric coupling across the interfaces.

\begin{figure}[t]\vspace{-0pt}
\includegraphics[width=7.5cm]{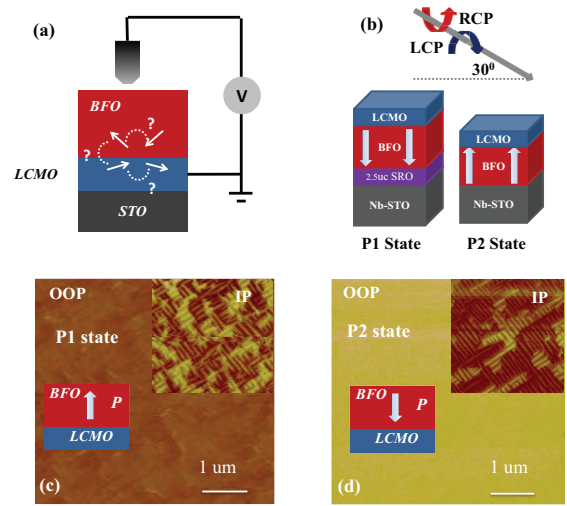}
\caption{\label{Schematic}  Schematic of the heterostructures: (a) structure H1 for ferroelectric switch, (b) structure H2 for XAS, XMCD and XLD; (c),(d) PFM of H1 structure in P1 and P2 states.}
\end{figure}

One model heteroepitaxial interface is comprised of the FM La$_{0.7}$Sr$_{0.3}$MnO$_{3}$ (LSMO) juxtaposed with the multiferroic BiFeO$_{3}$ (BFO). Previous studies on this model heterostructure have reported the emergence of enhanced, interfacial ferromagnetism in BFO and the possibility for interfacial orbital reconstruction in the manganite\cite{Yu0}. Furthermore, it was shown that the FE polarization controls the magnetization and magnetic anisotropy of FM manganite\cite{Yu1}. Within the broader framework of electric field control of magnetism, an interesting question can be put forward: besides controlling the magnetization direction and magnetic anisotropy of the ferromagnet, can one reversibly switch an antiferromagnet (with no macroscopic magnetic moment) to a ferromagnet (with a macroscopically sensible moment)? This is the central focus of this paper.

There are some logical criteria that can be used as “design rules” to accomplish this. First of all, it is desirable in these materials that ferromagnetism strongly competes with antiferromagnetic (AFM) ordering. Moreover, the competition should depend on external fields, such as electric/magnetic field, chemical potential or strain\cite{Tokura,Cui,Ryan}. One ideal candidate is the half-doped manganite, exemplified by La$_{0.5}$Ca$_{0.5}$MnO$_{3}$ (LCMO). Bulk LCMO undergoes two successive transitions: paramagnetic to FM transition followed by FM to AFM transition\cite{Radaelli}. Moreover, the transition is controllable by the magnetic field. These facts clearly characterize the strong competition between FM and AFM order. Recently Yin et al observed an electrically controllable tunneling resistance by inserting a thin (1-5 uc) LCMO barrier in the junction of LSMO/BTO/LSMO\cite{Yin}. While it suggests intriguing possibility to ferroelectrically induce a metal-insulator phase transition in LCMO, direct evidence for a ferroelectrically controllable AFM-FM transition has not be reported.

Here we report the ferroelectric control of AFM-FM transition at the heterointerface between LCMO and BFO. The large FE polarization of BFO provides a possible pathway to reversibly control the magnetic coupling of both the LCMO and BFO (schematic in Figure~\ref{Schematic}(a)). The combination of magnetometry and x-ray magnetic circular dichroism (XMCD) shows that the magnetization of both LCMO and BFO at the interface are modulated significantly through the reversal of the FE polarization. X-ray absorption spectra (XAS) at oxygen \emph{K}-edge and linear dichroism (XLD) at Mn \emph{L}-edge suggest that the main interfacial coupling is derived from the AFM superexchange between Fe and Mn \emph{t}$_{2g}$ spins. Furthermore, XAS reveals a modulation of the Mn 3\emph{d} electron occupancy due to the charge accumulation. Mean-field theory calculations suggest that the occupancy of Mn \emph{e}$_{g}$ electrons controls the magnetic coupling, thus tuning the competition between FM and AFM instability.


LCMO/BFO (5 nm/100 nm) heterostructures (Structure H1,Figure~\ref{Schematic}(a)) were prepared on TiO$_{2}$-terminated STO (001) substrate by pulsed laser deposition. Reflection high-energy electron diffraction (RHEED) was employed to achieve the atomic scale control of the heterointerface. An atomically-flat interface was observed by high-resolution scanning transmission electron microscopy (Figure S1\cite{Sup}). The magnetization of the heterostructures was measured by SQUID magnetometry. XAS was acquired by recording the total electron yield (TEY) current as a function of x-ray photon energy at the Beamline 4.0.2 of the Advanced Light Source at Lawrence Berkeley National Laboratory. Taking into account the surface-sensitivity (with a 30$^{o}$ grazing angle) and element-selectivity of TEY mode, XMCD was used to probe the element-resolved magnetic moments in the heterostructure. XLD was measured to study the orbital occupancy.


The FE polarization states of BFO were studied by the piezoresponse force microscopy (PFM). A metal-probe set up with a 50 $\mu$m tip was used to switch the polarization of the whole sample (Figure~\ref{Schematic}(a)). The dark contrast of the out-of-plane (OOP) image indicates that polarization points away from the interface (P1 state, Figure~\ref{Schematic}(c)) and the light contrast corresponds to the polarization towards the interface (P2 state, Figure~\ref{Schematic}(d)). PFM taken at multiple randomly selected regions confirms a nearly 100\% polarization control.


\begin{figure}[t]\vspace{-0pt}
\centering
\includegraphics[width=8cm]{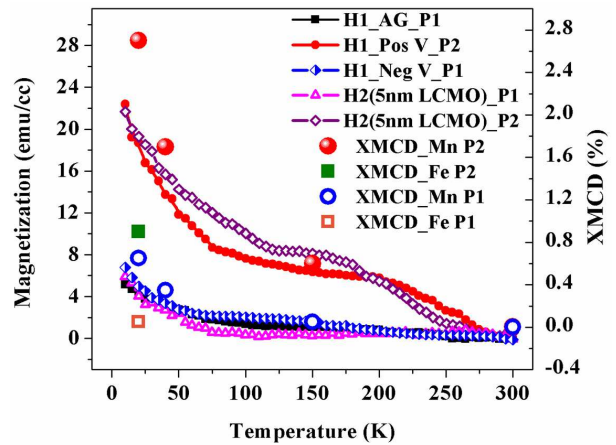}
\caption{\label{Squid}  Temperature dependence of magnetization of P1 and P2 states taken with magnetometer and calculated with spin sum-rule of XMCD.}
\end{figure}

By virtue of the well-controlled FE polarization, we measured the resulting change of the macroscopic magnetization. In Figure~\ref{Squid}, the black, red and blue curve shows the magnetization of the heterostructure (H1,figure ~\ref{Schematic}(a)) in the as-grown state (P1), switched to the opposite state (P2) by positive voltage and switched back to the original state (P1) by negative voltage. Reference data from a sample with 100 nm BFO on STO, which accounts for the diamagnetic signal of substrate and the small bulk canted moments of BFO\cite{Edere,Bea}, was subtracted from the raw data (figure S3 \cite{Sup}). The temperature dependence of magnetization in P1 state shows a negligibly small moment. On the other hand, magnetization of P2 state shows a macroscopically sensible moment. The saturation-like behavior between 100K and 200K suggests the FM clusters or canted AFM ordering\cite{Solovyev0,Kawano}. Furthermore, the data clearly shows that the modulation effect is reversible through FE switch. In order to identify the contribution from LCMO and BFO respectively, we utilized the element-specific XAS and XMCD to study Fe and Mn \emph{L}-edge.

\begin{figure*}[t]\vspace{-0pt}
\includegraphics[width=17cm]{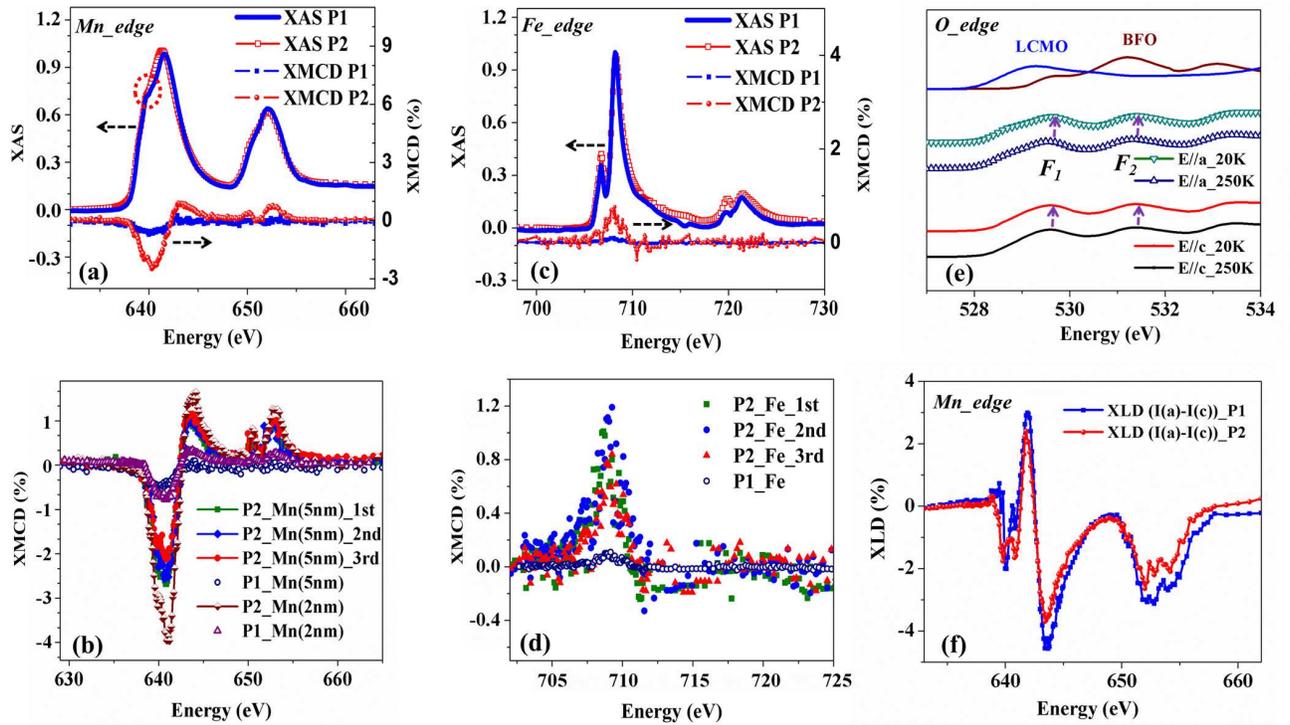}
\caption{\label{XAS} (a),(c) The XAS and XMCD spectra of Mn, Fe \emph{L}$_{2,3}$ edge in P1 and P2 states; (b),(d) multiple repeats of XMCD of Mn, Fe \emph{L}$_{2,3}$ edge; (e) XAS of oxygen \emph{K}-edge with linear polarized x-ray; (f) XLD spectra of Mn \emph{L}$_{2,3}$ edge (I(a) and I(c) correspond to XAS measured with in-plane and out-of-plane polarization). XLD is measured at 300K where the magnetic dichroic effect is absent. All the data are taken from H2 (5nm LCMO) except (b).}
\end{figure*}

Figure~\ref{XAS} (a)-(d) display the XAS and XMCD at the Mn and Fe \emph{L}-edge at 20K with an applied field of 0.5T. Due to the limited probing depth, the heterostructure with 100nm BFO top layer (H1) is not suitable to study the interface. Therefore, we grew the reversed heterostructure (H2, Figure~\ref{Schematic}(b)) of the BFO (100nm, bottom) and LCMO (5nm/2nm, top). In H2 structure, the TEY signal comes from the entire LCMO layer and the interfacial BFO. Our previous study demonstrated that the FE polarization can be controlled through the electrostatic boundary condition in the as-grown state\cite{Yu2}, which is also confirmed in the H2 structures (figure S2\cite{Sup}). Thus we can study the XAS and XMCD in the two FE states, i.e. P1 and P2. Magnetization taken with magnetometer shows the same modulation effects in H2 structure (5nm LCMO) (Figure~\ref{Squid}). The XMCD of $\sim$2.6\% at Mn \emph{L}-edge is clearly observed in P2 state (figure~\ref{XAS}(a)), which was confirmed in multiple samples (figure~\ref{XAS}(b)). However, no clear XMCD is observed in the P1 state. In addition, XMCD of Fe \emph{L}-edge also reveals a significant change($\sim$1\% in P2 state and negligible in P1 state, figure~\ref{XAS}(c) and (d)). The opposite sign of XMCD signifies an AFM coupling between Fe and Mn across the interface. The temperature dependence of XMCD is summarized in Figure~\ref{Squid}. The magnitude of the magnetization can be quantitatively estimated by the XMCD spin sum-rule (figure S4\cite{Sup}). The obtained value is roughly 0.25 $\mu$$_{b}$/Mn for LCMO and 0.1 $\mu$$_{b}$/Fe for BFO in P2 state at 20K. Considering the AFM coupling of Fe and Mn across interface, the calculated value is in reasonable agreement with the SQUID magnetometry. Furthermore, the XMCD of H2 structure with 2nm LCMO ($\sim$4\%) is larger than that of H2 structure with 5nm LCMO (figure~\ref{XAS}(b)), which suggests that the enhanced moments are mainly from interface. Interestingly, despite of the clear magnetization modulation, the transport property does not display significant variation for P1 and P2 state (figure S5(b)), unlike that of the tunneling structure studied by Yin et al\cite{Yin}.

Although the enhanced canted moment of BFO has been studied in LSMO/BFO interface\cite{Yu0}, the electrically controllable Fe edge XMCD has not been reported before. In order to gain further insight, we studied the microscopic magnetic coupling across the Fe-O-Mn bond. Figure~\ref{XAS}(e) shows the XAS of oxygen \emph{K} edge by using linear polarized x-ray at different temperatures (similar for P1 and P2, figure S6\cite{Sup}). The feature F1 corresponds to the mixture of Fe (\emph{t}$_{2g}$) and Mn (\emph{t}$_{2g}$ and \emph{e}$_{g}$) states and F2 is related to only the \emph{e}$_{g}$ levels of BFO. Previous study\cite{Yu0} observed that F2 peak shifts to lower energy as the temperature decreases in LSMO/BFO when the x-ray polarization is out-of-plane, which is explained as the hybridization of Mn and Fe 3z$^{2}$-r$^{2}$ orbital at the interface. However this shift is absent in Figure~\ref{XAS}(e). In contrast to LSMO, LCMO is under stronger tensile strain from the substrate, resulting in the stabilization of x$^{2}$-y$^{2}$ orbital compared with 3z$^{2}$-r$^{2}$, which is supported by the negative sign of XLD (I(a)-I(c)) at Mn \emph{L}-edge\cite{Pesquera} (Figure~\ref{XAS}(f)).  Therefore, the orbital reconstruction proposed for BFO/LSMO interface is not expected here. Instead, we speculate that the main magnetic coupling is derived from the AFM SE between Fe and Mn \emph{t}$_{2g}$ spins.

Based on the element-specific technique above, we found that the modulation effect is derived from both LCMO and BFO. We speculate that the FE polarization is likely to favor the FM (AFM) coupling in LCMO in P2 (P1) state due to the change of \emph{d}-electron density\cite{Vaz}. The varied magnetic coupling in LCMO then leads to the change of canted moments in the interfacial BFO due to the magnetic coupling. Besides the carrier modulation, strain effect should also be considered. The strain controlled nonvolatile magnetoelectric coupling requires the change of FE domain and thus the in-plane lattice constant\cite{Zhang}. However both P1 and P2 states in this study show similar four variant domains, which suggests that strain is not likely to be the main reason for the observed nonvolatile magnetization modulation.

To test the electronic origin, we performed a close examination on the XAS spectra of Mn \emph{L}-edge in the two polarization states (Figure~\ref{XAS}(a)). Although the XAS of Mn \emph{L} edge is similar to a +3/+4 mixed valence in both cases, there are a few clear differences between the two states. In particular, the XAS spectrum of P1 state (blue line) shows an enhanced shoulder-like feature on the low-energy side of the main peak of the \emph{L}$_{3}$ edge, which is distinctly absent in the spectrum of P2 state (red line with square symbol). Besides, the main absorption peak shifts to the higher energy level in P1 state comparing with the P2 state by roughly 0.2 eV. Previous studies on manganite revealed that both the peak position and the line shape of Mn \emph{L} edge XAS are highly sensitive to the Mn valence state\cite{Lee,Takamura}. It has been demonstrated that the peak energy increases for higher oxidation state of Mn, and the shoulder-like multiplet structure of the \emph{L}$_{3}$ edge is the fingerprint of Mn$^{4+}$ state. Moreover, the change of \emph{L}$_{3}$/\emph{L}$_{2}$ ratio follows the trend demonstrated in previous studies\cite{Varela}, which suggests higher oxidation state in P1 state. Therefore, by taking all these observations into account, we can reach a conclusion that the valence state of Mn changes due to the carrier modulation by the FE polarization. The valence state of P1 is closer to the Mn$^{4+}$, while the valence state of P2 is driven toward the Mn$^{3+}$. Based on the energy shift\cite{Cramer}, we estimate an average change of Mn valence to be $\sim$0.1/Mn, consistent with the calculated average change of charge of 0.11\emph{e}/Mn, assuming 2P$_{s}$=130 $\mu$C/cm$^{2}$\cite{Li}.

\begin{figure*}[t]\vspace{-0pt}
\includegraphics[width=17cm]{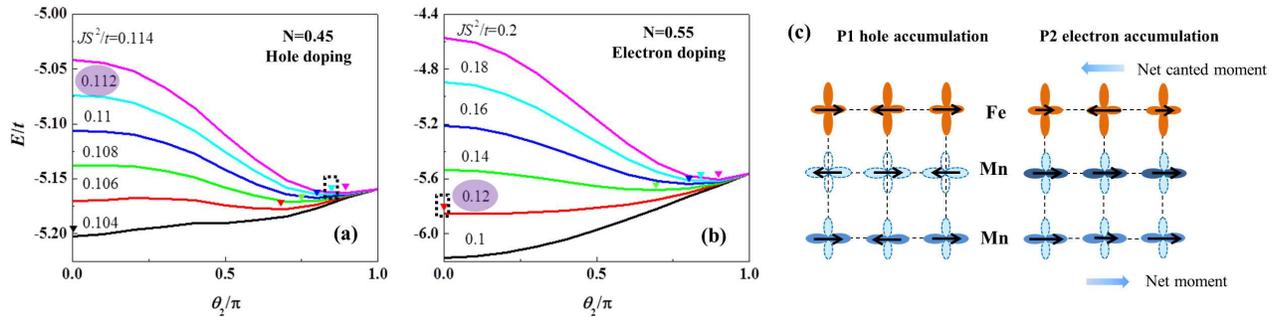}
\caption{\label{MFT}  Mean-field theory calculations on the relationship between \emph{e}$_{g}$ electron density (N) and spin structure for hole (a) and electron (b) doping of LCMO ($\theta$$_{2}$ is the relative angle between neighboring spins across the zigzag chain); (c) schematic of the electron density and spin structure at the heterointerfaces (The dark/light contrast suggests the electron accumulation/depletion).}
\end{figure*}


To quantify the relationship between the densities of \emph{e}$_{g}$ electrons and the magnetic interactions in LCMO, we then performed mean-field theory calculations. Here we consider a 2-dimensional double-exchange (DE) model with \emph{e}$_{g}$ orbitals and superexchange (SE) between neighboring \emph{t}$_{2g}$ spins. The details of the model are described in the supplementary materials (Figure S7(a)\cite{Sup}). The competition between SE coupling (JS$^{2}$) and DE (t) is represented by their ratio JS$^{2}$/t. For each value of JS$^{2}$/t, the relative angle $\theta$$_{2}$ between neighboring spins across the zigzag chain was obtained by solving for the minimum of free energy. ($\theta$$_{2}$=0/$\pi$ corresponds to the FM/CE-type AFM spins arrangement).  We estimate the critical value of JS$^{2}$/t to be slightly larger than 0.112, which reproduces the CE-AFM ground state with orbital ordering for N=0.5 (figure S7(b)\cite{Sup}), consistent with previous studies\cite{Solovyev1,Brink,Okamato}. Figure~\ref{MFT}(a) and~\ref{MFT}(b) present the results for hole doping (N=0.45) and electron doping (N=0.55) respectively. The results show that canting of antiferromagnetically coupled neighboring moments is possible for both electron and hole doping. However the canting is stronger in the electron-doping than the hole-doping side. The results suggest that a FM ordering is energetically more favorable at the critical value of JS$^{2}$/t when electrons are accumulated in LCMO, which is in accordance with the macroscopic moments observed by both the magnetometry and XMCD in P2 state. On the other hand, the AFM coupling is energetically more favorable in the hole-doping side. Based on these considerations, we propose a mechanism for the electrical control of magnetic coupling at BFO/LCMO heterointerface as shown schematically in figure~\ref{MFT}(c). The FM interaction in LCMO is enhanced by electron doping. The magnetic coupling across the heterointerface then leads to the larger canted moments in BFO. Oppositely both LCMO and BFO remain AFM in the hole-doping side. Our microscopic mechanism consistently explains the modulated competition between FM and AFM instability by switching FE polarization.


To summarize, we have demonstrated the ability to modulate the competition between FM and AFM instability through magnetoelectric coupling at BFO/LCMO heterointerface. The magnitude of magnetization in both the interfacial BFO and LCMO changes dramatically in response to the FE polarization switch. The magnetoelectric coupling is derived from the charge modulation and the interplay between charge and spin degree of freedom both in layer and across the interface. Our results suggest a possible route to explore the reversible electrical switch between an antiferromagnet and a ferromagnet. Indeed, we believe that there may be other similar pathways by which an AFM state can be reversibly switched into a FM state, through local electronic structure modulations.

The authors acknowledge fruitful discussions with G. Bhalla, J.H. Chu and G. Palsson. Research at Berkeley was sponsored by the National Science Foundation through the Penn State MRSEC. Work at ORNL was supported by the US Department of Energy, Basic Energy Sciences, Materials Sciences and Engineering Division. Work in National Chiao Tung University was supported by the National Science Council, R.O.C. (NSC-101–2119-M-009–003-MY2), Ministry of Education, R.O.C. (MOE-ATU 101W961), and Center for interdisciplinary science of National Chiao Tung University.


\begin{thebibliography}{100}
\bibitem{Hwang}
H. Y. Hwang et al., Nature Mater. \textbf{11}, 103 (2012).

\bibitem{Zubko}
P. Zubko et al., Annu. Rev. Condens.Matter Phys. \textbf{2}, 141 (2011).

\bibitem{Chakhalian}
J. Chakhalian et al., Nature Mater. \textbf{11}, 92 (2012).

\bibitem{Eerenstein}
W. Eerenstein et al., Nature \textbf{442}, 759 (2006).

\bibitem{Cheong}
S. W. Cheong and M. Mostovoy, Nature Mater. \textbf{6}, 13 (2007).

\bibitem{Ramesh}
R. Ramesh and N. A. Spaldin, Nature Mater. \textbf{6}, 21 (2007).

\bibitem{Fiebig}
M. Fiebig, J. Phys. D  \textbf{38}, R1239 (2005).

\bibitem{Vaz}
C. Vaz et al., Phys. Rev. Lett. \textbf{104}, 127202 (2010).

\bibitem{Burton}
J. Burton and E. Tsymbal, Phys. Rev. B \textbf{80}, 174406 (2009).

\bibitem{Wu0}
S. M. Wu et al., Nature Mater. \textbf{9}, 756 (2010).

\bibitem{Wu1}
S. M. Wu et al., Phys. Rev. Lett. \textbf{110}, 067202 (2013).

\bibitem{Yu0}
P. Yu et al., Phys. Rev. Lett. \textbf{105}, 027201 (2010).

\bibitem{Yu1}
P. Yu et al., unpublished work.

\bibitem{Tokura}
Y. Tokura, Rep. Prog. Phys. \textbf{69}, 797 (2006).

\bibitem{Cui}
C. Cui et al., Phys. Rev. B \textbf{68}, 214417 (2003).

\bibitem{Ryan}
P. Ryan et al., Nature Comm. \textbf{4}, 1334 (2012).

\bibitem{Radaelli}
P. Radaelli et al., Phys. Rev. B \textbf{55}, 3015 (1997).

\bibitem{Yin}
Y. W. Yin et al., Nature Mater. \textbf{12}, 397 (2013).

\bibitem{Sup}
Supplementary information.

\bibitem{Edere}
C. Ederer and N. Spaldin, Phys. Rev. B \textbf{71}, 060401(R) (2005).

\bibitem{Bea}
H. Béa et al., Appl. Phys. Lett. \textbf{87}, 072508 (2005).

\bibitem{Solovyev0}
I. Solovyev et al.,Phys. Rev. B \textbf{63}, 174425 (2001).

\bibitem{Kawano}
H. Kawano, Phys. Rev. B \textbf{53}, 2202 (1996).

\bibitem{Yu2}
P. Yu et al., PNAS \textbf{109}, 9710 (2012).

\bibitem{Pesquera}
D. Pesquera et al., Nature Communication \textbf{3}, 1189 (2012).

\bibitem{Zhang}
S. Zhang et al., Phys. Rev. Lett. \textbf{108}, 137203 (2012).

\bibitem{Lee}
J. Lee et al.,Phys. Rev. B \textbf{80}, 205112 (2009).

\bibitem{Takamura}
Y. Takamura et al.,Phys. Rev. B \textbf{80}, 180417(R) (2009).

\bibitem{Varela}
M. Varela et al., Phys. Rev. B \textbf{79}, 085117 (2009).

\bibitem{Cramer}
S. P. Cramer et al., J. Am. Chem. Soc. \textbf{113}, 7937 (1991).

\bibitem{Li}
J. Li et al., Appl. Phys. Lett. \textbf{84}, 5261 (2004).

\bibitem{Solovyev1}
I. Solovyev et al., Phys. Rev. Lett. \textbf{83}, 2825 (1999).

\bibitem{Brink}
J. Brink et al., Phys. Rev. Lett. \textbf{83}, 5118 (1999).

\bibitem{Okamato}
S. Okamoto, Phys. Rev. B \textbf{82}, 024427 (2010).







\end{thebibliography}
\end{document}